\documentstyle[12pt]{article}
\makeatletter
\@addtoreset{equation}{section}
\makeatother

\begin{document}
\begin{center}
{\Large \bf
On Stapp-Unruh-Mermin Discussion on Quantum
Nonlocality:\\[0.5cm]
Quantum Jumps and Relativity}
\\[1.5cm]
{\bf Vladimir S.~MASHKEVICH}\footnote {Email:
mash@gluk.apc.org}  \\[1.4cm]
{\it Institute of Physics, National academy
of sciences of Ukraine \\
252028 Kiev, Ukraine} \\[1.4cm]
\vskip 1cm

{\large \bf Abstract}
\end{center}

We argue that the participants of the discussion have overlooked
an essential circumstance, in view of which Stapp's fifth
proposition fails. The circumstance is that though $L$ and $R$
measurements, being causally separated, are not invariantly
time ordered, quantum-jump hypersurfaces associated with the
measurements are causelikewise ordered. Stapp's fifth
proposition is true iff $L$ hypersurface precedes $R$ one; but
within the limits of special relativity, it is impossible to
determine the causelike order of those hypersurfaces. The
entire Stapp's construct is revised.

\newpage

\section*{Introduction}

In a recent paper [1] Stapp made a sophisticated attempt to
breathe new life into the problem of quantum nonlocality. The
paper raised objections by Unruh [2] and Mermin [3], which
resulted in a discussion on that problem [4-8].

The aim of Stapp's construct is the conclusion that there exists
a backward-in-time influence. Stapp's fifth proposition is the
cornerstone of the construct. Unruh and Mermin argue that the
conclusion is untenable, though they grant the fifth
proposition. Finkelstein [8] argues that the fifth proposition
should be specified through consideration of a hypothetical
world.

We argue that the participants of the discussion have
overlooked an essential circumstance, in view of which Stapp's
fifth proposition fails.

The circumstance is that though $L$ and $R$ measurements [1-8],
being causally separated from each other, are not
Lorentz-invariantly ordered in time, hypersurfaces of quantum
jumps associated with the measurements are causelikewise
ordered. Stapp's fifth proposition is true if and only if the
$L$-hypersurface precedes the $R$-hypersurface. But within the
limits of special relativity, it is impossible to determine---
both theoretically and experimentally---what are those
hypersurfaces and which is their causelike order.

Thus, the entire Stapp's construct should be revised, which
is done in the present paper.

Invoking reference frames and time order for causally separated
events is misleading, so that we use the geometric, or
intrinsic approach to the problem.

\section{Geometric, or intrinsic description of
quantum field}

\subsection {Coordinate-free description}

To avoid questions concerning reference frames, we shall be
based on a geometric, or intrinsic description of a quantum
field.

In the Heisenberg picture, a quantum field is
\begin{equation}
\phi_{H}\equiv \phi=\phi(p),\quad p\in M,
\label{1.1}
\end{equation}
where $M$ is the Minkowski spacetime,
\begin{equation}
\Psi_{H}\equiv\Psi
\label{1.2}
\end{equation}
is a state vector;
\begin{equation}
\phi^{{\rm class}}(p)=(\Psi,\phi(p)\Psi)
\label{1.3}
\end{equation}
is a classical field.

\subsection{Coordinate, or reference-frame description}

Let $\{\tilde x\}$ and $\{\bar x\}$ be two Lorentzian
coordinate systems,
\begin{equation}
p\leftrightarrow\tilde x\leftrightarrow\bar{x},
\qquad \bar{x}=\Lambda\tilde{x}+a.
\label{1.4}
\end{equation}
We have
\begin{equation}
\phi(p)\leftrightarrow\{\tilde{\phi_{i}}(\tilde{x}):i\in
{\cal I}\}
\leftrightarrow\{\bar{\phi_{i}}(\bar{x}):i\in {\cal I}\}.
\label{1.5}
\end{equation}
$\Psi$ is the same both in the geometric and in all coordinate
descriptions.

The transformation
\begin{equation}
\bar{\phi}^{{\rm class}}_{i}(\bar{x})=S_{ij}(\Lambda)
\tilde{\phi}^{{\rm class}}_{j}(\tilde{x})
\label{1.6}
\end{equation}
implies
\begin{equation}
\bar{\phi}_{i}(\bar{x})=S_{ij}(\Lambda)\tilde{\phi}_{j}
(\tilde{x}).
\label{1.7}
\end{equation}

Now we have
\begin{equation}
S_{ij}(\Lambda)\tilde{\phi}_{j}(\tilde{x})=U^{-1}(a,\Lambda)
\tilde{\phi}_{i}(\tilde{x}')U(a,\Lambda)
\label{1.8}
\end{equation}
where
\begin{equation}
\tilde{x}\leftrightarrow p\ne p'
\leftrightarrow \tilde{x}'=
\Lambda\tilde{x}+a=\bar{x}.
\label{1.9}
\end{equation}

Let us take $\{\tilde{x}\}$ as a fiducial coordinate system and
put
\begin{equation}
\tilde{\Psi}=\Psi,
\label{1.10}
\end{equation}
then
\begin{equation}
(\tilde{\Psi},\bar{\phi}_{i}(\bar{x})\tilde{\Psi})=
(\bar{\Psi},\tilde{\phi}_{i}(\tilde{x}')\bar{\Psi})
\label{1.11}
\end{equation}
where
\begin{equation}
\bar{\Psi}=U(a,\Lambda)\tilde{\Psi}.
\label{1.12}
\end{equation}

\section{Quantum jumps: The problem of appropriate
hypersurfaces and causelike ordering}

In the Heisenberg picture, a state vector $\Psi$ changes at
and only at quantum jumps. To every quantum jump there
corresponds a spacelike hypersurface. The hypersurfaces must
be mutually disjoint: otherwise $\Psi$ would be not defined.
The quantum-jump hypersurfaces are causelikewise ordered:
\begin{equation}
{\cal S}_{2}>{\cal S}_{1},\qquad {\rm or} \;\;{\cal S}_{1}<
{\cal S}_{2}.
\label{2.1}
\end{equation}
Within the limits of special relativity, it is impossible to
determine---both theoretically and experimentally---what are
those hypersurfaces and which is, in the general case, their
causelike order (see Section 4 below). It is special
relativity that prevents a complete phenomenological
mathematical description of quantum jumps and specifically
measurements.

(A complete dynamical description of quantum jumps has
been given in the
series of our papers [9], which is beyond the scope of the
present paper.)

\section{Two-particle system}

\subsection{$LR$-system}

Following [1] and the discussion, we consider a two-particle
$LR$-system. Let
\begin{equation}
\rho_{LR}\equiv\rho
\label{3.1}
\end{equation}
be a statistical operator of the system in the Heisenberg
picture. We have
\begin{equation}
\rho_{R}={\rm Tr}_{L}\rho,\quad \rho_{L}={\rm Tr}_{R}\rho.
\label{3.2}
\end{equation}

\subsection{Local measurements and associated quantum jumps}

A measurement implies instantaneity, which---in the limits of
special relativity---makes sense locally only. A local
measurement for $L$ or $R$ system results in an associated
quantum jump for the state of $LR$ system:
\begin{equation}
\{{\cal L}\}\Rightarrow\rho^{0}\stackrel{\{{\cal L}\}}
{\longrightarrow}
\rho^{\{{\cal L}\}},\quad \{{\cal R}\}\Rightarrow\rho^{0}
\stackrel{\{{\cal R}\}}{\longrightarrow}\rho^{\{{\cal R}\}}
\label{3.3}
\end{equation}
where $\{{\cal L}\}$ stands for measuring an observable
${\cal L}$ of
$L$ system.

Let two measurements be performed: $\{{\cal L}\}$ at a point
$p_{\{{\cal L}\}}\in M$ and $\{{\cal R}\}$ at a point
$p_{\{{\cal R}\}}
\in M$;
let ${\cal S}_{\{{\cal L}\}}$ and ${\cal S}_{\{{\cal R}\}}$ be
hypersurfaces of the associated quantum jumps. We write
\begin{equation}
p_{\{{\cal L}\}}<(>)p_{\{{\cal R}\}}\quad {\rm iff}\quad
{\cal S}_{\{{\cal L}\}}<(>)
{\cal S}_{\{{\cal R}\}}.
\label{3.4}
\end{equation}

Let ${\cal L}$ and ${\cal R}$ be integrals of motion. We have
\begin{equation}
{\rm for}\quad p_{\{{\cal L}\}}<p_{\{{\cal R}\}}:\quad \rho^{0}\stackrel
{\{{\cal L}\}}
{\longrightarrow}\rho^{\{{\cal L}\}}\stackrel{\{{\cal R}\}}
{\longrightarrow}\rho^{\{{\cal L}{\cal R}\}},
\label{3.5}
\end{equation}
\begin{equation}
{\rm for}\quad p_{\{{\cal R}\}}<p_{\{{\cal L}\}}:
\quad \rho^{0}\stackrel{\{{\cal R}\}}
{\longrightarrow}\rho^{\{{\cal R}\}}\stackrel{\{{\cal L}\}}
{\longrightarrow}\rho^{\{{\cal R}{\cal L}\}};
\label{3.6}
\end{equation}
the equality
\begin{equation}
\rho^{\{{\cal L}{\cal R}\}}=\rho^{\{{\cal R}{\cal L}\}}
\label{3.7}
\end{equation}
holds since
\begin{equation}
[{\cal L},{\cal R}]=0.
\label{3.8}
\end{equation}

Let $p_{\{{\cal L}\}}$ and $p_{\{{\cal R}\}}$ be causally
separated, then
there exist frames $\{\tilde{x}\}$ and $\{\bar{x}\}$, such
that
\begin{equation}
\tilde{t}_{\{{\cal L}\}}<\tilde{t}_{\{{\cal R}\}},\quad \bar{t}_
{\{{\cal L}\}}>
\bar{t}_{\{{\cal R}\}};
\label{3.9}
\end{equation}
thus, we must not appeal to the time order.

Within the limits of special relativity, it is impossible
to determine which of the two cases (\ref{3.5}),(\ref{3.6})
takes place (see Section 4 below).

\subsection{Nonselective and selective measurements}

It is necessary to discriminate between nonselective and
selective measurements [10]: in a (non)selective measurement
the result is (not) registered.

Let
\begin{equation}
{\cal L}\left| l \right\rangle=l\left| l \right\rangle,\qquad
\left\langle l \right| l \rangle=1.
\label{3.10}
\end{equation}

A nonselective measurement $[{\cal L}]$ is described by
\begin{equation}
\rho^{0}\stackrel{[{\cal L}]}{\longrightarrow}\rho^{[{\cal L}]}=
\sum_{l}\left| l \right\rangle\left\langle l \right|\rho^{0}
\left| l \right\rangle\left\langle l \right|,\quad
\rho^{[{\cal L}]}\ne\rho^{0},
\label{3.11}
\end{equation}
\begin{equation}
\rho_{L}^{0}={\rm Tr}_{R}\rho^{0}\stackrel{[{\cal L}]}
{\longrightarrow}
\rho_{L}^{[{\cal L}]}={\rm Tr}_{R}\rho^{[{\cal L}]}=
\sum_{l}\left| l \right\rangle\left\langle l \right|\rho_{L}^{0}
\left| l \right\rangle\left\langle l \right|,\quad
\rho_{L}^{[{\cal L}]}\ne\rho_{L}^{0},
\label{3.12}
\end{equation}
\begin{equation}
\rho^{0}_{R}={\rm Tr}_{L}\rho^{0}\stackrel{[{\cal L}]}
{\longrightarrow}
\rho_{R}^{[{\cal L}]}={\rm Tr}_{L}\rho^{[{\cal L}]}=\sum_{l}
\left\langle l \right|\rho^{0}\left| l \right\rangle=
{\rm Tr}_{L}\rho^{0},\quad \rho_{R}^{[{\cal L}]}=\rho^{0}_{R}.
\label{3.13}
\end{equation}

A selective measurement $(l)$ is defined by
\begin{equation}
\rho^{0}\stackrel{(l)}{\longrightarrow}\rho^{(l)}=
\frac{\left| l \right\rangle\left\langle l \right|
\rho^{0}\left| l \right\rangle\left\langle l \right|}
{{\rm Tr}\{\left| l \right\rangle\left\langle l \right|
\rho^{0}\left| l \right\rangle\left\langle l \right|\}}=
\frac{\left| l \right\rangle\left\langle l \right|
\rho^{0}\left| l \right\rangle\left\langle l \right|}
{\left\langle l \right|\rho^{0}_{L}\left| l \right\rangle},
\quad \rho^{(l)}\ne\rho^{0},
\label{3.14}
\end{equation}
\begin{equation}
\rho_{L}^{0}\stackrel{(l)}{\longrightarrow}\rho_{L}^{(l)}=
{\rm Tr}_{R}\rho^{(l)}=\left| l \right\rangle\left\langle l
 \right|,\quad \rho^{(l)}_{L}\ne\rho_{L}^{0},
\label{3.15}
\end{equation}
\begin{equation}
\rho^{0}_{R}\stackrel{(l)}{\longrightarrow}\rho_{R}^{(l)}=
{\rm Tr}_{L}\rho^{(l)}=\frac{\left\langle l \right|
\rho^{0}\left| l \right\rangle}
{\left\langle l \right|\rho^{0}_{L}\left| l \right\rangle},
\quad \rho_{R}^{(l)}\ne\rho_{R}^{0}.
\label{3.16}
\end{equation}

\subsection{Pure initial state; symmetric case and reciprocity
relation}

Let
\begin{equation}
\rho^{0}=\left| 0 \right\rangle\left\langle 0 \right|.
\label{3.17}
\end{equation}

A selective measurement $(l)$ is described by
\begin{equation}
\left| 0 \right\rangle\stackrel{(l)}{\longrightarrow}
\left| (l) \right\rangle=\frac{\left| l \right\rangle
\left\langle l \right|\left. 0 \right\rangle}
{\|\left| l \right\rangle\left\langle l \right|
\left. 0 \right\rangle\|}=
\frac{\left| l \right\rangle\left\langle l \right|
\left. 0 \right\rangle}
{\|\left\langle l \right|\left. 0 \right\rangle\|},
\label{3.18}
\end{equation}
so that
\begin{equation}
\left| (l) \right\rangle=\left| l \right\rangle
\left| R(l) \right\rangle,\quad \left| R(l) \right\rangle=
\frac{\left\langle l \right|\left. 0 \right\rangle}
{\|\left\langle l \right|\left. 0 \right\rangle\|}.
\label{3.19}
\end{equation}

The symmetric case is that where
\begin{equation}
\frac{\left\langle R(l) \right|\left. 0 \right\rangle}
{\|\left\langle R(l) \right|\left. 0 \right\rangle\|}=
\left| l \right\rangle.
\label{3.20}
\end{equation}

Let a selective measurement for $L$ result in $\left| L
\right\rangle$ and
\begin{equation}
\left\langle L \right|\left. 0 \right\rangle\propto
\left| R \right\rangle,
\label{3.21}
\end{equation}
so that
\begin{equation}
\left| L \right\rangle\Rightarrow\left| R \right\rangle.
\label{3.22}
\end{equation}
From
\begin{equation}
\left\langle \perp R \right|\left. R \right\rangle=0
\label{3.23}
\end{equation}
it follows
\begin{equation}
\left\langle \perp R \right|\left\langle L \right|\left.
 0 \right\rangle=0,\quad \left\langle L \right|
\left\langle \perp R \right|\left. 0 \right\rangle=0,
\quad \left\langle \perp R \right|\left. 0 \right\rangle
\propto \left| \perp L \right\rangle,
\label{3.24}
\end{equation}
so that
\begin{equation}
\left| \perp R \right\rangle \Rightarrow
\left|\perp L \right\rangle.
\label{3.25}
\end{equation}
Thus we have a reciprocity relation:
\begin{equation}
(\left| L \right\rangle\Rightarrow\left| R \right\rangle)
\Rightarrow(\left| \perp R \right\rangle\Rightarrow
\left| \perp L \right\rangle).
\label{3.26}
\end{equation}

\subsection{Actual measurements: Locality or nonlocality?}

Here the question relates to actual measurements. Nonlocality
is an influence of an $\{{\cal L}\}$ measurement at
$p_{\{{\cal L}\}}$ on
an $R$-state at $p_{R}$, with $p_{\{{\cal L}\}}$  and $p_{R}$ being
causally
separated. In view of eqs. (\ref{3.13}),(\ref{3.16}), the
influence exists if and only if the measurement is selective.
Every actual measurement is selective, so that quantum theory
is nonlocal.

We quote Bell [11]: ``The paradox of Einstein, Podolsky and
Rosen ... was advanced as an argument that quantum mechanics
could not be a complete theory but should be supplemented by
additional variables. These additional variables were to
restore to the theory causality and locality...''

It seems pertinent to quote Unruh [2] as well: ``... locality is
usually used to argue that value that a variable attains must
be independent of the choice of experiment carried out in a
causally disconnected region (although correlations clearly mean
that the value need not be independent of the values obtained
for measurements in disconnected regions).''

\section{Impossibility of determining causelike order of jumps
caused by causally separated measurements}

Let $(l)$ and $(r)$ be selective measurements of ${\cal L}$ and
${\cal R}$
respectively, $p_{(l)}$ and $p_{(r)}$ being causally separated.
Let us find a conditional probability $P(l\mid r)$ for
$p_{(r)}<p_{(l)}$ and $p_{(l)}<p_{(r)}$. Let $N$ be the number of
measurements. We have
\begin{equation}
{\rm for}\quad  p_{(r)}<p_{(l)}:\quad N=N_{r}+N_{\bar{r}}=(N_{rl}+
N_{r\bar{l}})+N_{\bar{r}}
\label{4.1}
\end{equation}
where $\bar{r}$ stands for not $r$, so that
\begin{equation}
P_{rl}(l\mid r)=\frac{N_{rl}}{N_{r}}=\frac{N_{rl}/N}{N_{r}/N}=
\frac{P(rl)}{P(r)};
\label{4.2}
\end{equation}
\begin{equation}
{\rm for}\quad p_{(l)}<p_{(r)}:\quad N=N_{l}+N_{\bar{l}}=
(N_{lr}+N_{l\bar{r}})+(N_{\bar{l}r}+N_{\bar{l}\bar{r}})=
(N_{lr}+N_{\bar{l}r})+(N_{l\bar{r}}+N_{\bar{l}\bar{r}}),
\label{4.3}
\end{equation}
\begin{equation}
P_{lr}(l\mid r)=\frac{N_{lr}}{N_{lr}+N_{\bar{l}r}}.
\label{4.4}
\end{equation}

Now we have
\begin{equation}
N_{lr}=N_{rl}\quad {\rm since} \quad [{\cal L},{\cal R}]=0
\label{4.5}
\end{equation}
and
\begin{equation}
N_{lr}+N_{\bar{l}r}=N_{r}=N_{rl}+N_{r\bar{l}}\quad {\rm since}\quad
\rho_{R}^{[{\cal L}]}=\rho_{R}^{0}.
\label{4.6}
\end{equation}
Thus
\begin{equation}
P_{lr}(l\mid r)=P_{rl}(l\mid r).
\label{4.7}
\end{equation}

\section{Counterfactuals}

\subsection{Counterfactual probabilities}

Following [1] and the discussion, we consider counterfactuals.
Let ${\cal R}'$ be an observable for $R$, such that
\begin{equation}
[{\cal R}',{\cal R}]\ne 0.
\label{5.1}
\end{equation}
Let $[{\cal L}]$ and $(r)$ measurements be performed,
$p_{[{\cal L}]}$ and
$p_{(r)}$ being causally separated. Our interest here is with
the counterfactual probability
\begin{equation}
P^{{\rm cf}}(r')\equiv P^{{\rm cf}}((r')\mid \{[{\cal L}],(r)\}).
\label{5.2}
\end{equation}
$P^{{\rm cf}}(r')$ depends on a state $\rho_{R}$ in which
${\cal R}'$ would be measured and $r'$ would be obtained:
\begin{equation}
P^{{\rm cf}}(r')=P(r'\mid \rho_{R})=\left\langle r' \right|
\rho_{R}\left| r' \right\rangle.
\label{5.3}
\end{equation}
We have
\begin{equation}
{\rm for} \quad p_{(r)}<p_{[{\cal L}]}:\quad \rho_{R}=\rho_{R}^{0},
\label{5.4}
\end{equation}
so that
\begin{equation}
P^{{\rm cf}}_{r{\cal L}}(r')={\rm Tr}_{L}\left\langle r' \right|
\rho^{0}\left| r' \right\rangle;
\label{5.5}
\end{equation}
\begin{equation}
{\rm for}\quad p_{[{\cal L}]}<p_{(r)}:\quad \rho_{R}=\sum_{l}
P_{{\cal L}r}(l\mid r)\rho_{R}^{(l)},
\label{5.6}
\end{equation}
or, in view of eqs. (\ref{4.7}) and (\ref{3.16}),
\begin{equation}
\rho_{R}=\sum_{l}P_{r{\cal L}}(l\mid r)\frac{\left\langle l
 \right|\rho^{0}\left| l \right\rangle}
{\left\langle l \right|\rho^{0}_{L}\left| l \right\rangle}.
\label{5.7}
\end{equation}
Now
\begin{equation}
P_{r{\cal L}}(l\mid r)=\frac{P(rl)}{P(r)}=
\frac{\left\langle l \right|\left\langle r \right|
\rho^{0}\left| r \right\rangle\left| l \right\rangle}
{\left\langle r \right|\rho^{0}_{R}\left| r \right\rangle},
\label{5.8}
\end{equation}
so that
\begin{equation}
P^{{\rm cf}}_{{\cal L}r}(r')=\sum_{l}\frac{\left\langle r \right|
\left\langle l \right|\rho^{0}\left| l \right\rangle
\left| r \right\rangle\left\langle r' \right|
\left\langle l \right|\rho^{0}\left| l \right\rangle
\left| r' \right\rangle}{\left\langle r \right|
\rho^{0}_{R}\left| r \right\rangle\left\langle l \right|
\rho^{0}_{L}\left| l \right\rangle}.
\label{5.9}
\end{equation}
We emphasize that
\begin{equation}
P^{{\rm cf}}_{r{\cal L}}(r')\ne P^{{\rm cf}}_{{\cal L}r}(r'),
\label{5.10}
\end{equation}
which undermines Stapp's fifth proposition.

\subsection{Pure initial state}

For $\rho^{0}$ given by eq. (\ref{3.17}), we have
\begin{equation}
P^{{\rm cf}}_{r{\cal L}}(r')={\rm Tr}_{L}\{\left\langle r'
 \right|\left. 0 \right\rangle\left\langle 0 \right|
\left. r' \right\rangle\},
\label{5.11}
\end{equation}
\begin{equation}
P^{{\rm cf}}_{{\cal L}r}(r')=\sum_{l}\frac{\mid
\left\langle r \right|\left\langle l \right|
\left. 0 \right\rangle\left\langle r' \right|
\left\langle l \right|\left. 0 \right\rangle\mid^{2}}
{{\rm Tr}_{L}\{\left\langle r \right|\left. 0 \right\rangle
\left\langle 0 \right|\left. r \right\rangle\}
{\rm Tr}_{R}\{\left\langle l \right|\left. 0 \right\rangle
\left\langle 0 \right|\left. l \right\rangle\}}.
\label{5.12}
\end{equation}

\subsection{Two-dimensional Hilbert spaces}

For two-dimensional Hilbert spaces of $L$ and $R$ systems,
we have
\begin{equation}
\sum_{l}f(l)=f(l)+f(\bar{l}),\quad \sum_{r}g(r)=g(r)+
g(\bar{r}).
\label{5.13}
\end{equation}

\section{Hardy-type initial state}

Following [1] and the discussion, let us consider a Hardy-type
initial state (though all is already done by eq. (\ref{5.10})):
\begin{equation}
\left| 0 \right\rangle=\frac{\Psi}{\|\Psi\|},\quad  \left| \Psi
 \right\rangle=\left| l' \right\rangle\left| r' \right\rangle
-\langle \bar{l} \mid l' \rangle
\left\langle r \right|\left| r' \right\rangle
\mid \bar{l} \rangle\left| r \right\rangle
\label{6.1}
\end{equation}
where
\begin{equation}
{\cal L}=L2,\;{\cal L}'=L1,\;{\cal R}=R2,\;{\cal R}'=R1;
\qquad l=L2+,\;\bar{l}=L2-,\;
l'=L1+,\;r=R2+,\;r'=R1-.
\label{6.2}
\end{equation}
We find
\begin{equation}
{\rm for}\quad p_{(r)}<p_{[{\cal L}]}:\quad P^{{\rm cf}}
_{r{\cal L}}(r')=
\frac{\mid\left\langle l \right|\left. l' \right\rangle\mid
^{2}+\mid\langle \bar{l} \mid l' \rangle
\mid^{2}(1-\mid\left\langle r \right|\left. r'
 \right\rangle\mid^{2})^{2}}
{\mid\left\langle l \right|\left. l' \right\rangle\mid^{2}
+\mid\langle \bar{l} \mid l' \rangle\mid
^{2}(1-\mid\langle r \mid r' \rangle\mid^{2})}
<1,
\label{6.3}
\end{equation}
\begin{equation}
{\rm for}\quad p_{[{\cal L}]}<p_{(r)}:\quad P^{{\rm cf}}
_{{\cal L}r}(r')=1.
\label{6.4}
\end{equation}
It is impossible to determine which case---(\ref{6.3}) or
(\ref{6.4})---takes place. If $p_{(r)}<p_{[{\cal L}]}$,
the $[{\cal L}]$-measurement has nothing to do with $P^{{\rm cf}}
_{r{\cal L}}
(r')$. If and only if $p_{[{\cal L}]}<p_{(r)}$, the (nonselective)
$[{\cal L}]$-measurement allows for nontrivial statements on
counterfactual measurements for $R$.

\section{On the Stapp-Unruh-Mermin discussion}

Stapp's fifth proposition [1] is the cornerstone of Stapp's
construct. Unruh and Mermin argue that the conclusion of the
construct is untenable, but they grant the fifth proposition.
Finkelstein [8] argues that the fifth proposition should be
specified through consideration of a hypothetical world.

We argue that Stapp's fifth proposition is incorrect. Let us
consider that proposition and its proof in the form given by
Mermin [3]:

``($I$) Whenever the choice of measurement on the left is $L2$,
if the measurement on the right is $R2$ and gives $+$, then
if $R1$ were instead performed the result would be $-$.

... The validity of this is established by translating it into
the language appropriate to the frame of reference in which
the events on the left happen first:

($I_{L}$) Whenever the choice of measurement on the left was
$L2$, if the measurement done later on the right is $R2$ and
gives $+$, then if $R1$ were instead done later on the right,
the result would have to be $-$.''

In ($I_{L}$) a frame is used in which
\begin{equation}
t_{[{\cal L}]}<t_{(r)}.
\label{7.1}
\end{equation}
Had eq. (\ref{7.1}) implied $p_{[{\cal L}]}<p_{(r)}$,
\begin{equation}
t_{[{\cal L}]}<t_{(r)}\Rightarrow p_{[{\cal L}]}<p_{(r)},
\label{7.2}
\end{equation}
($I_{L}$) would have followed from eq. (\ref{6.4}). But the
implication (\ref{7.2}) is wrong; indeed eq. (\ref{7.1}) has
nothing to do with the problem of nonlocality. It is misleading
to invoke reference frames and a time order for causally
separated events.

\section*{Conclusion}

Quantum theory is nonlocal. As for actual measurements,
nonlocality manifests itself in and only in selective
ones. As for counterfactual nonselective measurements
$[{\cal L}]$, nonlocality manifests itself if and only if
$p_{[{\cal L}]}<p_{(r)}$, the fulfillment of which cannot be
established in the limits of special relativity.

Quantum jumps and special relativity are incompatible:
the former cannot be described in the limits of the latter.

\section*{Acknowledgment}

I would like to thank Stefan V. Mashkevich for helpful
discussions.

\end{document}